\documentclass[aps,twocolumn,groupedaddress, showpacs,prb]{revtex4}
\usepackage{graphicx}%

\begin{document}
\bibliographystyle{apsrev}


\title{A simplex model for layered niche networks}


\author{P. Fraundorf}
\affiliation{Physics \& Astronomy, U. Missouri-StL (63121), St. Louis, MO, USA}
\email[]{pfraundorf@umsl.edu}


\date{\today}

\begin{abstract}

The standing crop of correlations in metazoan 
communities may be assessed by an 
inventory of niche structures focused inward and outward 
from the physical boundaries of skin (self), gene-pool
(family), and meme-pool (culture).  We consider
tracking the progression from three and four correlation layers
in many animal communities, to five of six layers for the 
shared adaptation of most humans, with an attention-slice 
model that maps the niche-layer focus of individuals 
onto the 6-variable space of a 5-simplex.  
The measure puts questions about the effect, on culture 
and species, of policy and natural events 
into a common context, and may help explore 
the impact of electronically-mediated codes on 
community health.

\end{abstract}
\maketitle

\tableofcontents
\section{Introduction}

Thinking on multiple scales of space and time 
is a crucial element of the move toward 
sustainability.  Here we discuss 
quantification of a multi-scale view of organization 
among metazoan individuals, which draws from 
roots in physical representation theory \cite{Lapilli2006}.  
This may be useful, for 
example, as we explore the short-term impact of new 
technologies as well as the long-term window of 
opportunity for metazoan life \cite{Ward03}.

The approach here builds on recent findings in three convergent 
disciplines.  The first of these is nanoscience, where 
chemistry, physics, biology, engineering, medicine, 
crime scene investigation, ethics, and complex system 
studies of emergence run together.  The second is 
astrobiology, where Eric Chaisson's {\em Cosmic 
Evolution}\cite{Chaisson04, Chaisson98} (a natural history 
of invention with physical science roots) and 
Brownlee and Ward's {\em Rare Earth}\cite{Ward00} intersect 
our distant past, and our distant future.  The third 
is complex system informatics, where the code-based sciences 
(genetics, computer science, linquistics), thermal 
physics, journalism, networks and statistical inference 
join up.  For the most part, however, independent 
of their technical context the ideas described 
should also be possible to check against the reader's 
everyday experience.  Thus we won't
belabor the technical roots except when necessary.

\section{Correlation layers and boundaries}

\begin{figure}
\includegraphics[scale=.8]{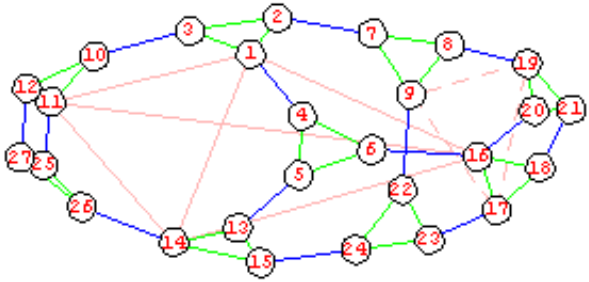}%
\caption{Layered niche network inventory: This one shows 
four scales of correlation: self (\#), friends (blue), 
family (green) \& teams (light red).}
\label{Fig1}
\end{figure}

One way to inventory social complexity is to think of it as a layered network of niches occupied by individuals, as illustrated in Fig. \ref{Fig1}.  Here we define {\em niche} as a correlation through time between one steady-state system (e.g. an organism or heat engine), or more generally one excitation (e.g. a particle), and its environment.  Niche network inventories in this sense can be applied to complex systems on layers ranging from elementary (e.g. particle state inventories in statistical physics) to complex (e.g. metazoan communities). Of course in the more complex cases: Operational definitions, ennumeration of alternate states, and objective implementation may be quite challenging, particularly if one wishes to put the resulting measures into $2^{nd}$ Law terms.  


\begin{figure}
\includegraphics[scale=.5]{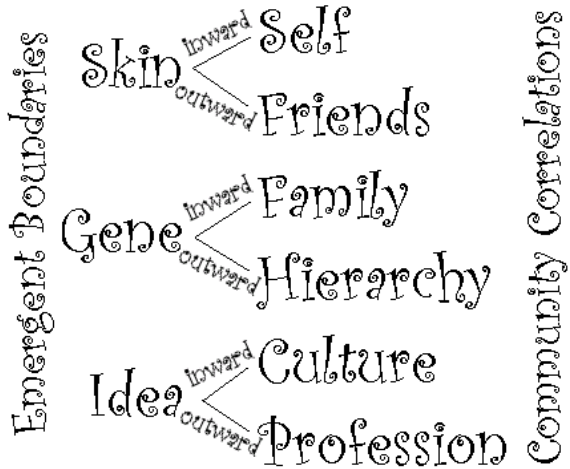}%
\caption{Community correlation schematic}
\label{Fig2}
\end{figure}


Returning to the community focus of this paper, the layers of a 
network built up from 
the level of individual metazoans are often conceptualized as 
inward- and outward- looking with respect
to three increasingly larger physical boundaries, 
namely your skin, your family's gene 
pool, and your culture's idea pool.  The pattern is 
illustrated in Figure \ref{Fig2}.  In other words, our social
fabric is built on everyone's ability to take care of themselves, 
and be responsible to their friends, family, job, culture, and 
knowledge base.

The extent to which citizens can fill 
niches on all six of these levels is a measure of community 
complexity.  The impact of policy decisions and 
disasters on community health may also be reflected by 
changes in the average number of niche levels (0-6) that
each citizen can participate in.  Increasing literacy, 
for example, likely improves a person's ability to 
contribute to the knowledge base with their own
observations.  Conversely, loss of one's home to a flood 
might wreak havoc on long term friendships and jobs, even 
if everyone's health and family stays intact.

Although an objective inventory of 
such niche occupancies is difficult (i.e. it is not easy to 
take the essence of a community and project it onto a spreadsheet), 
the spirit of a program which seeks to give everyone a chance 
to simultaneously take care of self, friends, family, hierarchy, culture and 
profession is well-grounded in our knowledge of simpler 
(e.g. physical) systems.  It may also be satisfying in other ways. 
Thus for the sake of future discussion, we consider the ability of 
members to occupy niches 
on all six scales of organization as a measure of community health.

\section{Simplex models of attention or resource focus}
\label{section3}

\begin{figure}
\includegraphics[scale=0.5]{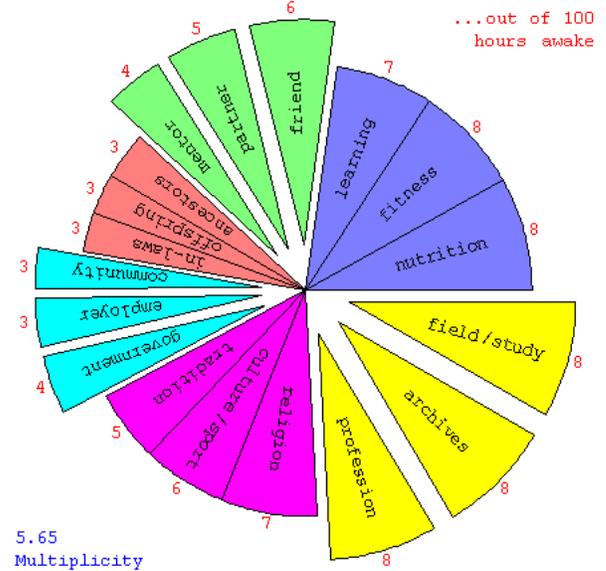}%
\caption{Attention fractions for a hypothetical individual}
\label{pieslices}
\end{figure}

To quantify, one might develop technical definitions for 
the above concepts (e.g. friendship as a pair correlation, 
political consensus as multi-family agreement, scholarly 
reports as codes that mirror nature, etc.).  These technical
definitions (like that for work in physics) could serve
as a complement to their more diverse vernacular forms.  
From that point, one might then associate with 
each of $i=1,N$ individuals that fraction $f_{ij}$
of their attention allocated to maintaining correlations 
on each of the j=1,6 niche scales, as 
discussed in Appendix \ref{AppxA}.  
For example, someone who spends directs all of 
their energy toward church-activities or dance-class 
(i.e. developing 
cultural connections) might have a larger value of 
$f_{i5}$ than someone who spends all their 
time doing applied mathematics.  The six $f$-values 
for an individual thus describe the distribution 
of size scales on which they connect to the
community around (cf. Fig \ref{pieslices}).

In addition to the distribution of individuals 
in this 6-parameter space, the average 
surprisal $s$ (in bits) of an individual's focus given only 
the $f_{ij}$ is one measure of level diversity for 
a community as a whole.  {\em Niche level multiplicity} defined 
as $2^s$, a number between 0 and the active 
number of niche layers, allows us to put this  
measure of community health into even simpler terms.

\begin{figure}
\includegraphics[scale=0.85]{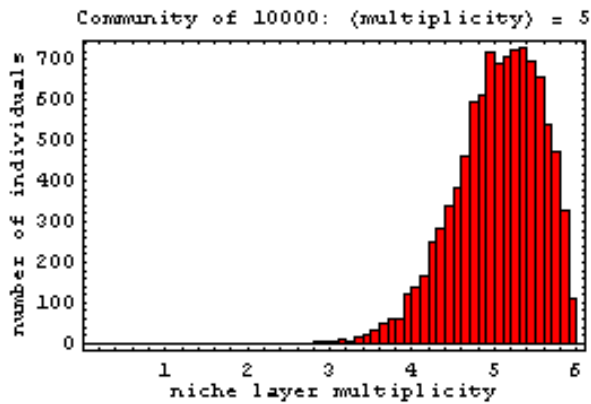}%
\caption{Distribution of niche level multiplicity for community of 10000}
\label{Histogram}
\end{figure}

The distribution of niche-level multiplicity for 
{\em individuals} in a community of 10000 is illustrated 
in Fig. \ref{Histogram}.  Here the attention fractions 
for each individual were chosen by assigning a 
random number between 0 and 1 to each, and then 
normalizing the set by dividing by the sum.  
This {\em ad hoc} protocol increases the likelyhood 
of small f-values over that of a uniform assignment 
in f-space.  It's as if each niche layer were assigned a 
stand-alone priority, later put into context by 
the fixed amount of time in a day. 

The resulting niche-level multiplicity of the community 
as a whole (a geometric average of individual 
multiplicities) in Fig. \ref{Histogram} is five out of six.  Potentially 
interesting questions include the following:  
In which column of this plot would you fit?  How will your 
niche level multiplicity change during the next decade?  How would  
similar plots look for the various communities of which you 
are a part?  How will changes in the environment and/or economy 
affect the profile for those communities?  

Since the six f-values for an individual are positive 
numbers which sum to one, they span the 
hypervolume of a unit-vertex 5-simplex in f-value's 
6D parameter space, as discussed in 
Appendix \ref{AppxB}.  Ternary sums which 
span this space (like $f_{i1}+f_{i2}+f_{i3}$, $f_{i4}$, $f_{i5}+f_{i6}$) 
thus plot as points in an equilateral triangle, while 
quaternary spanning sums (like 
$f_{i1}$, $f_{i2}$, $f_{i3}$, $f_{i4}+f_{i5}+f_{i6}$) 
plot as points in an equilateral tetrahedron.  
Individual states, without sum degeneracy, can also 
be plotted as {\em point pairs} in a set of vertex-joined 
tetrahedra (cf. Fig. \ref{FigX}).  Properties of 
simplex models applied to other natural systems \cite{Ohtsuki2006} 
may prove useful here as well.

\subsection{Advertising Example}

The niche layer for self-care might be used to distinguish 
folks who select coffee, tea, or neither when given a choice.
One could experimentally model the effect of ads on (i)
the probability $f_{i1}$ that folks will concentrate on self-care, 
and (ii) in so doing choose either coffee or tea.  Thus 
we are looking at niche occupancies, and their correlation 
with specific idea sets.  Advertisers are already 
doing this, even if not in this multiscale context.

\subsection{Highway Safety Example}

To maximize safety, automobile drivers 
should probably divide their attention between keeping their own 
vehicle safe, and not endangering others on the roads around.  
Simultaneous focus on other niche-level responsibilities while driving in 
this context serves as a distraction.  
Such distractions can move into real time when one is 
also holding and 
talking on a cell phone.  Measurements of the resulting 
attention loss might be 
put onto a more quantitative footing using the conceptual tools
discussed here.

\subsection{Applications in Social Psychology}

Control systems theory in 
sociology looks at the way that humans interface 
to the world through niche-related 
ideas.  Although these ideas differ from one culture 
to the next, the multiscale or layered-network 
structure discussed here offers an integrative thread.
For example, all human cultures distinguish
between family and non-family individuals even 
if the word associations used are different.  
Via connections like this, data generated e.g. 
by affect control experiments in social 
psychology \cite{PMA2006} may be used to monitor the effect 
of media events or social policy on objective 
measures of community health.  If we focus on 
the way that behavioral resources (rather than 
perceptual attention) are directed toward 
community correlations, this same integrative strategy 
may also be put to use in studying the behavioral ecology of 
animal communities centered more around genetically 
rather than culturally-transmitted codes.

\subsection{Political Science Example}

Here, let's select the social hierarchy layer and ask 
about the effect of ads on (i) the probability $f_{i4}$ that 
folks will take political responsibility in a given 
election, and (ii) in so doing choose one party or 
another.  Simulations over many decades here likely 
show some interesting oscillations.  Again media is already 
being put to use by politicians, and so a developing 
awareness of the effect of those actions on community 
health may be a good thing.

\section{Hypotheses to test}

Given a measure of community 
health (niche level multiplicity) referenced to 
three physical boundaries 
(skin, family gene pool, and cultural idea pool), 
we now consider hypotheses that data on 
this measure might allow us to test.
The focus will be on applications in 
human communities, and in particular 
the relationship of these measures 
to ideas (i.e. concept sets in use) and
their means of replication.

The relationship between ideas and niche focus is
of practical interest for two reasons.  First, 
along with behavioral observations \cite{Reeve01} and 
population surveys, 
communication traffic may provide important clues 
to the niche focus across a community.  Secondly, 
mediated communications have become increasingly important 
as a mechanism for manipulating the public's attention, 
particularly by those with commercial and/or political 
interests.  Multiscale thinking, with deep roots in the 
natural sciences, could offer a much needed public 
health perspective on these activities.

\subsection{Link to memetic evolution}

\begin{figure}
\includegraphics[scale=0.85]{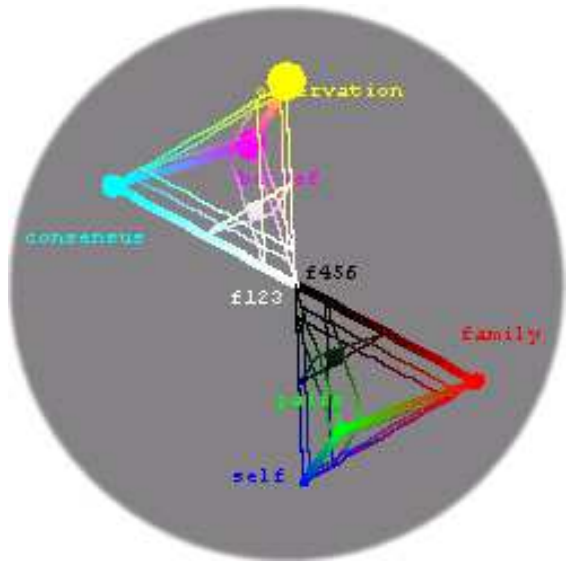}%
\caption{Dual simplex representation (cf. Appendix \ref{AppxB}) 
of a (1/6, 1/6, 1/6, 1/6, 1/6, 1/6) niche assignment.}
\label{FigX}
\end{figure}

Before man's development of written communication, 
{\em the separation} between one's cultural knowledge 
and one's
knowledge of nature's behavior was probably less well 
defined.  Local culture was the source of both 
types of knowledge \cite{Diamond92}.  In communities of 
animals with less developed oral communication skills,
cultural training itself seems to be less developed
but certainly not absent \cite{vanSchaik2003}.  
Thus it may be that the pre-history shared by all 
humans involved adaptation of individuals to five,
and not six, layers of niche development in parallel with our
shared refinement of oral communication skills. 

This is important, because 
adaptation to five niche levels may 
be something most humans have in common.  As humans face more 
complex challenges to their large population in days ahead, understanding 
what we can and cannot count on may be important.  Advertisers and 
politicians are already putting their knowledge of in-born 
human responses to work on us.  

Before discussing possible impacts of these 
in-born adaptations on community health, it's 
worth pointing out the recent philosophical interest 
in the ``selfish'' evolution of cultural codes \cite{Distin05}.  
The integrative context discussed here can 
provide reality checks from the physical sciences 
(e.g. the non-locality of mutual information 
in physical systems is not inherent in vernacular 
uses of the concept), and as we'll see 
may also connect it (ala McLuhan\cite{McLuhan62}) 
quantitatively to technology 
through representational systems running from 
language through print to the internet.  
The context of idea technology
naturally, then, leads us to questions about  
the opportunistic role of human-developed codes as an 
antagonist to robust niche development, and following 
that to prospects for giving as many humans 
as possible a chance to fill niches on all six scales 
of organization.

\subsection{Codes that redirect behavior}

By associating replicable codes with elements 
of our environment (including other organisms), 
ideas serve to redirect human behavior much 
as greeting behavior in animals serves to redirect 
intra-species aggression \cite{Lorenz66}.   
However, in addition to helping us maintain 
levels of organization in human communities, 
un-informed ``reduction to code'' can also cause trouble.  This is 
especially true now that ideas can be replicated electronically 
across the planet's surface in seconds, by anyone 
with access to a web cafe.  Put another way, 
self-consciousness about our choice of ideas may 
be crucial.  This is especially true of news 
media, when they insist that someone answer a 
particular question without justifying 
why that particular question is appropriate.

Here we discuss 
hypotheses associated with technological 
redirection of niche-forming behaviors.  These 
include the web's ability 
to amplify a neolithic survival trait (inter-cultural 
xenophobia) into oscillations of destructive 
cartoonification, and more generally 
ways to minimize the bad effects of opportunities 
created as technology evolves.

\subsubsection{Mediated interaction as niche substitute}

If one recognizes money as a 
technology developed along with 
the development of food production and distribution, then one can
also see that one of our oldest ``professions'', prostitution, 
arose as a way to technologically fake some 
rewarding aspects of a reproductive partnership with help from 
a once new technology (namely, the coin of the realm).  
Any technology that redirects behavior away 
from responsible interaction with other humans plays 
the same role.  

Just as money has its good side and its bad, so 
newer technologies (like video games) capable of distracting you from 
human interaction likely have positive uses e.g. 
in education, or as occasion for mental, physical, or 
team skills exercise.  Hypothesis one, then, is this:  
Emphasizing the importance of responsible 
niche participation on each of the levels discussed above might 
help folks keep their eyes on the positive (rather than 
the ``fake niche'') aspect of these technologies, 
and in the process increase rather than decrease 
niche multiplicity.

\subsubsection{Cultural cartoons: Xenophobia \& the web}

Our evolved response to the idea of a boogeyman, 
if we ignore the dynamics of this idea, allows it to run amok in the 
information age outside of video games as well as in.  For example
if you respond to someone calling you subhuman by calling them 
subhuman, then the idea of treating others as subhuman is given a 
boost by {\em both} 
of you whether or not that behavior is in anyone's interest.

\begin{figure}
\includegraphics[scale=0.85]{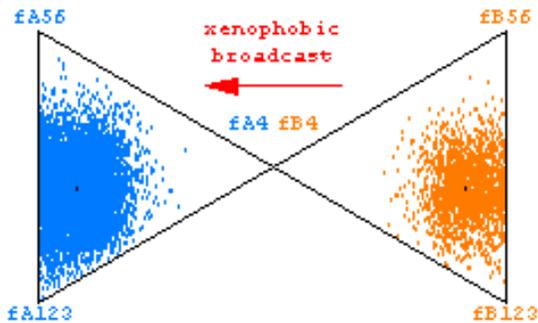}%
\caption{Mediated f-value interaction between two communities}
\label{Fig4}
\end{figure}

This is illustrated 
in the ternary plots of individual $f$-values for 
groups A and B in Fig. \ref{Fig4}, where a resource problem that 
might only be solved by scientific observation ($f_{i6}$) 
precipitates discomfort that results in a xenophobic 
broadcast from one group of 2000 individuals (B).   
Because in today's world that broadcast is accessible to all, 
this broadcast can shift the focus of both the 
uncomfortable group (B), and the object group (A) 
of 10,000 
individuals,
toward political action ($f_{i4}$). 

Electronic media capable of sharing broadcast and 
counterbroadcast 
may thus shift the focus of both populations away from 
where it is needed.
Note that movement of individual 
dots in the figure  
toward the $f_{i4}$ vertex denotes a shift of attention toward 
politics, regardless of the side they take on any given issue.     
If we recognize in this the dynamics of  
codes that had survival value long ago but can be improved 
upon now, then we have a better 
chance to keep such {\em oscillatory cartoonification} 
from causing more damage than necessary.  

Hypothesis two:  Such analyses might predict 
the toppling of each district's statues after 
deforestation on Easter Island \cite{Diamond04}, 
or the electronic ramping of  
intensity in current wars by humans not 
thinking about the dynamics of xenophobic ideas in a 
resource-limited world.  This is a special case 
of the ability of electronically-broadcast ideas to 
redirect the attention of large subsets of our 
population via evolved human 
perceptions.  To the extent 
that mirror neurons \cite{Rizzolatti2006} can 
directly elicit feelings via the faux intimacy 
of an electronic broadcast (as with the ``neighborhood
gossip'' ambience of some political attack ads), 
one can even imagine feedback loops that
manipulate the public's attention on visceral levels 
to bypass logic and self-interest 
altogether.

\subsection{Level devolution from 6 Down}

We discussed earlier how our evolved 
response to language likely involved mainly niches pertaining to 
self, friends, family, hierarchy, and culture.  This might help 
explain why some of us substitute our profession (if we find 
opportunity to develop one) for one of these other elements in 
our life.  


One practical consequence:  level blurring from 6 down 
to 5 increases the likelyhood of extremist niche occupancies, 
i.e. individuals whose behavior is informed to only a minor 
subset of the constraints 
associated with a 6-level niche structure.  For example, if 
one plots
contours of constant net 
surprisal (separated by 0.1 bit) with respect to an equal-fraction 
ambient (where all fractions are 1/N), when two levels become 
one those contours move outward toward the corners of the plot.
If extremist preoccupation (the ignoring of correlations 
outside of one corner) helps replicate an idea which drives this 
blurring process, such blurring could easily be amplified in this 
age of electronic communications {\em with no conscious input} on 
the part of the participants.    

When free energy per capita decreases, Eric 
Chaisson's anecdotal observations of the relation between complexity 
and free energy rate density suggest that complexity will also decrease,
and thus that pressure to devolve civilization from 6 back down to 
5 layers of individual correlation may increase.  We mention 
three examples of this here.

\subsubsection{Confusing observation \& consensus}

The scientific tradition involves 
observations of nature, faithfully reported, followed by the 
iterative refinement of conceptual models for its understanding
and prediction.  Social consensus about those models has its 
uses, but is generally a dated and belated spin off.  Scientific 
observation is the horse which pulls that cart.

Similarly teaching science is about 
training each of us as critical observer, not about teaching the 
consensus.  Thus when scientists and creationists argue about which 
consensus to teach, they are helping to blur the 
distinction between science and politics.  Their discussion 
generally does not involve what observational tools to use.  
Thus a mistaken idea of science as ``taught consensus'' 
has folks fighting for no good reason.  Once again our 
gut level reactions, in the face of ideas with electronic 
wings, are distracting us from the real challenges at hand.  
These challenges include decreasing free energy per capita 
that could help drive the confusion.

\subsubsection{Confusing consensus \& belief}

The separation between 
church and state is an old story, with new 
chapters being written every day.  We portray the drive to merge 
the two here in a different light, namely 
as community simplification (devolution) e.g. toward 
an endpoint where the alpha-wolf's majority decides on 
the ideas used.  In effect this leaves room for 
fewer cultural niches than there are citizens, and 
points toward a memetic monoculture with shortcomings 
(as far as survivability is concerned) similar to 
those of genetic monocultures in biology.  
On the contrary, layered niche models suggest that 
consideration of {\em endangered cultures} may be as important 
scientifically as consideration of endangered 
species, if only we can separate science, culture, 
and politics in addressing it.

\subsubsection{Confusing belief \& observation}

In the historical scheme of things the 
distinction between niches focussed on belief and observation
is probably most recent, and 
hence also the one we've had least time to evolve with.  As many
have said, of course, there is no {\em a priori} reason why 
belief and observation should be in conflict unless one 
seeks to promote beliefs that fly in the face of observation.  
The good news from the other direction is that 
this scientific approach to correlation-based complexity protects 
cultural beliefs as an element of structure as zealously as it does 
the fruits of scientific observation.  Simplifiers in either direction 
are where the real confusion lies.

\subsection{Level building toward 6}

We've talked a bit above about ideas 
which replicate easily among humans in an electronic age, but
that also break down community correlations in the process.  
We hypothesize here a few strategies that might help 
to strengthen correlations instead.

\subsubsection{Tracking soft correlations}

\begin{table*}
\caption{Boundaries for social organization in metazoan communities}
\begin{tabular}{c|c|c|c|c}
Inward Focus & Physical Boundary & Outward Focus & Applies To & Correlation Model  \\
\hline
Individual & Metazoan Skin & Env/Pairs/Friends & All & Patterns in SpaceTime \\
\hline
Family  &  Gene Pool  & Consensus Hierarchy & Animals & 4-scale resource slicing  \\
\hline
Culture/Beliefs & Meme Pool & Professional Observation & Idea Sharers & 6-scale attention slicing \\
\end{tabular}
\label{TableX}
\end{table*}

Rather than concentrate only on 
economic measures and population (which are relatively easy to 
be objective about), consider also trying to track 
niche-layer occupancy per citizen (0-6) in a community 
as a function of time.  Why?  One reason is that it 
at least tries to register the 
impact of policy changes on deeper aspects of community 
structure.  Although objective quantification 
complicated, since even the participants may not be sure how well 
they are doing, this is a place where technology might 
be applied to social health as it is already applied to 
the health of individuals.  As economists look for measures 
of progress linked to the steady state rather than to 
rates of growth, this might be useful there as well.

\subsubsection{Context on message scope}

Although it is flattering to be asked one's opinion 
about all manner of subjects, the fact is that we have 
a mandate and resources to be informed about some 
subjects more than others.  For example, 
a company manager's (or policy's) goals and effectiveness 
will look different from the vantage point of: (i) a 
company employee, (ii) one of the company's customers, 
and (iii) the company's board of directors.  That's 
because the information environment, as well as the 
mandate, of each of these parties is quite different.  
Each input may be relevant to informed evaluation.  
Similarly by using 
non-technical vernacular, science popularizers often ruffle 
the feathers of their more technically particular colleagues
even though they are much more fun for the general 
public to listen to.  

In both cases, tagging messages with information on 
(i) their source, and (ii) the intended scope of their 
audience, could help.  As with food tagged according 
to source and intended destination, and with political 
messages bundled with a candidate's admission that
they endorse it, such accompanying
information can mediate one's visceral reaction to the 
message (or meal) by prompting them to ask:  Is the 
sender informed to what I should be thinking 
about, and is my reaction to it part of a larger 
context that I want to support or not?  What layers of 
organization, and what populations, are being 
served thereby?  Such information puts messages into 
a multiscale context.

\subsubsection{Complementarity: Bundling code \& organism perspectives}

Get in the habit of talking about code 
plus organism perspectives when reporting news, and 
use idea codes for their modern relevance rather 
than only because they appeal to our prehistoric 
selves.  For example, when two people 
get into a fight after the exchange of ``inappropriate gestures''
toward one another, in addition to reporting the human spectacle the 
reporter might examine the history of such gestures 
(which in related form even occur among non-human primates in the wild).  
There are other places where organism behavior
is not the whole story.

The discussion of Fig. \ref{Fig4} is a 
case in point.  If the press in 
region B describes a protest group in region 
A as the enemy, an oscillation of
the sort mentioned above might start {\em in 
the absence of opposition} to begin with.  
That is, in this information environment 
an oscillation might nucleate from random 
fluctuations and be amplified into larger 
opposing pushes while the underlying
resource problem falls off the radar.  
Linking the politics ($f_{i4}$) to religion 
($f_{i5}$) or family ($f_{i3}$) might 
increase the amplitude of excursions toward the 
political corner of the niche-focus 
simplex, thus increasing extremism by 
lessening the pull toward other corners.  
Thus the media of this example might play 
an unwitting role via its 
critical focus on organisms but not ideas.

A global electronic media 
ignoring replicable code dynamics and limitations, but 
focused on pushing buttons as cheaply as possible, could 
thus be doing more harm than good.  This 
leads to the hypothesis that yellow 
journalism can negatively impact community health 
quantitatively, in terms of niche-layer multiplicity.

\subsubsection{Newsgroups vs. chats}

Finally, instant messaging and chat rooms naturally 
represent their participants as replicable
code, and have a reputation for eliciting ``flame wars'' 
between participants.  
These is another case of technologically redirected niche-forming 
behavior, since they take place in a world of replicable codes, and 
they can and are turned off at the flick of a switch much like a 
video game battle when it's time to go to bed.

However technology 
also has potential for archiving responsible interaction, and for 
saving correlations for access by future generations.  Usenet 
newsgroups, recently hooked into browsers and cloned in 
moderated-access form by Google, are a case in point.  Here, niches of 
accountable interaction on many levels can be developed,  
nurtured, and later studied.  Newsgroup work for over a decade by 
John Baez on current finds in mathematical physics is a case in point.  
These mediated streams allow one to iteratively develop 
healing ideas for a community of interest, and to simultaneously
document the process for participants downstream.

\section{Discussion}

Thus correlation inventories with roots in the physical 
sciences may prove useful in monitoring the state and 
process dynamics of layered niche networks, including the 
metazoan communities of which we are a part.  Other 
interesting features of this approach include the observation 
that idea sets (e.g. values) targeted toward the maintenance 
of specific levels of correlation in such inventories are 
easily identified, and that the stone age history that most 
humans share \cite{Diamond92} likely adapted us primarily to 
deal with only five of these six levels. 

We outline a simplex model based on the 
focus of attention among metazoans toward 
the buffering of community correlations 
looking inward and outward from the edges 
of body, family, and culture.  Anecdotal 
data, obtained largely from communication traffic 
particularly in human communities, suggests 
a number of interesting applications for 
this model.  However {\em quantitative}
behavioral observations, population 
surveys, and data on communication 
focus as well, are signficant hurdles at 
this point to putting the model to use.

\begin{acknowledgments}
I would like to thank Eric Chaisson at Tufts, Peter Rogan at UM-KC, Edwin Taylor at MIT, Tom Schneider of NIH, Don Brownlee at U. Washington, and the late E. T. Jaynes at Washington University for many interesting papers and discussions.
\end{acknowledgments}

\begin{appendix}
\section{Why these six, and only six, levels?}
\label{AppxA}

Recognizing correlations between 
systems on different size scales probably requires that 
we know where a given system starts and ends, or in other words 
a definition of boundaries.  In the case of complex 
systems, their boundaries generally start out as ill-defined but 
over time emerge as worth taking for granted.

Examples of this over 
time include the development of a star from a diffuse 
cloud of gas, the formation of planets from a bunch of 
tiny dust particles in orbit, the development of predictably 
structured enzyme-molecule surfaces from mixed atoms in 
a random soup, the formation of bilayer cell membranes with 
highly sophisticated pore structures to facilitate reproduction
of useful biochemical cycles, the development of skin-bound 
metazoans with fancy 
endocrine, nervous, gastro-intestinal, circulatory, and immune 
systems from simpler tissue alliances like those found in jelly 
fish, etc.  In each case, the new boundary arrives out of 
the blue and in poor focus.  For a finite time, however, 
these boundaries begin to take on clear and practical 
significance.  Each new boundary, in turn, is 
a jumping off point for emergence of the next 
boundary in our tree of layered correlations (cf. 
Table I here \cite{pf.roots}).

\begin{figure}
\includegraphics[scale=0.85]{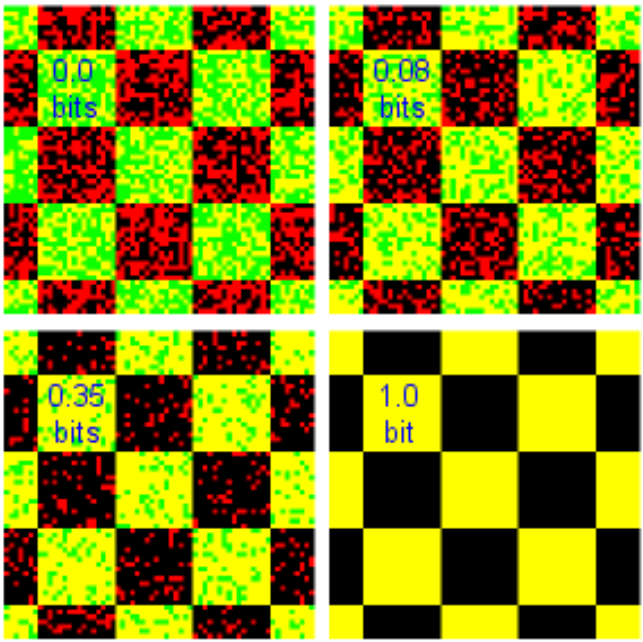}%
\caption{Geographic correlations in a population}
\label{FigA1}
\end{figure}

On size scales up 
from that of multi-celled organisms suppose that 
we consider only three such boundaries, namely 
metazoan skins, molecular code-pool boundaries 
(already complicated in a geometric sense), and 
idea code-pool boundaries (e.g. as physically encoded in 
recordable speech, writing, and a wide range of replicable 
digital formats including video).
At first glance it seems that plants correlate with 
their environment spatially (cf. Fig. \ref{FigA1}), but 
otherwise act more
as heat engines (which capture thermodynamic availability) 
than as information 
engines (which buffer the kinds of code-pool correlation
discussed here).  For example, what plants show bias
toward their own weaned offspring?  

\begin{figure}
\includegraphics[scale=0.85]{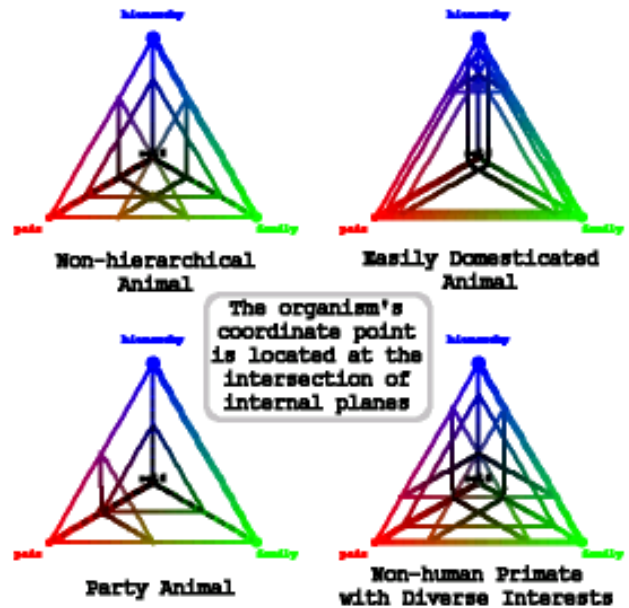}%
\caption{Limiting cases of resource-focus on 
the 3-simplex with foci looking inward and 
outward from skin (self-pair) and gene-pool 
(family-hierarchy).  The ``party animal'' 
assignment (sans negative connotation) in the lower 
left corner, namely {1/2,1/2,0,0}, refers to an 
organism invested in friendships and self, 
without regard to elements of the larger picture. 
It might also characterize youngsters of a species 
who've not yet shouldered the responsibilities of 
an adult.}
\label{FigA2}
\end{figure}

On the other hand, along with 
self-care animals seem comparatively active at pair interaction, 
as well as in the buffering of correlations directed inward (family) 
and outward (hierarchy) from their immediate gene pool.
Activity focused inward and outward from skin and gene 
pool can be represented by the tetrahedral 3-simplex 
shown in Fig. \ref{FigA2}. 
The investment of most animals in correlations 
directed inward and outward from  
cultural boundaries, however, seems pale in comparison 
to the attention directed by humans.  In order of 
increasing layer-scale the three boundaries (Table \ref{TableX}) 
thus give rise to six very popular subjects of human 
discourse, namely health care, 
pair bonds, family matters, political hierarchy, 
culture/religion, and scientific/extra-cultural lore.

In even more basic physical terms, 
the magic of complex 
systems comes from the correlations which make the 
whole more than a sum of parts.  This is the mathematical 
definition of mutual 
information.  Niches that target the
nurturing of correlations of the above 6 types (e.g. intra-family 
or inter-family) might comprise much of that magic in 
communities.  In principle one could directly 
measure the effect of behaviors on correlations of this 
sort, like the spatial correlations between red organisms and environmental 
greenery in the above figure at right (recall that red PLUS green 
equals yellow).  However, the dynamical nature of 
layered network correlations, as well as the  problems that 
plague measurement of 
algorithmic complexity in computer programs (e.g. we 
are likely to miss important emergent 
patterns), make this an incomplete solution.  We 
therefore suggest asking {\bf for} each individual: 
``What fraction of their attention and/or effort is 
targeted toward the buffering of subsystem 
correlations in each of these 6 areas?''   These are the 
$f$-values mentioned in Section \ref{section3}.

This ``attention-slice'' strategy for 
quantifying correlations in animal communities 
thus builds on recognition of two kinds of thermodynamic 
symbiosis.   
The first kind is the {\em informatic
symbiosis} between steady state excitations 
and replicable codes, integral 
to life of all sorts.  For example, 
microbes store information on how to 
make useful proteins in nucleic 
acid codes whose perspective is as 
important as that of the microbes 
that carry them.  Similarly, 
governmental hierarchies store information 
on useful procedures in books of law which 
carve out their own evolutionary tale.  

The second kind is the 
{\em trophic symbiosis} between multi-celled 
plants and animals, wherein the former (autotrophs) 
convert thermodynamic availability (for 
the most part from sunlight caught by 
leaves facing skyward) into available work which 
in turn is thermalized by the latter (heterotrophs)   
while constructing their layered networks of 
subsystem correlation.  The latter 
correlations include lively friendships, 
family activity, team accomplishments, 
cultural traditions, as well as  
libraries of observational data (with unavoidable
interpretation) on how the world 
around works.  No surprise then that 
absence of these traits is sometimes 
compared to a vegetable-like existence.

\section{Representations of the 5-simplex}
\label{AppxB}
\end{appendix}

\begin{figure}
\includegraphics[scale=0.85]{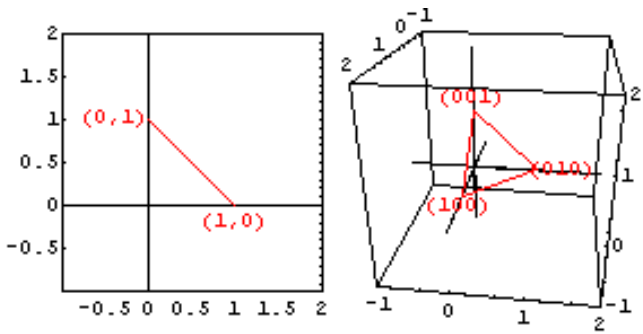}%
\caption{Simplex mapping of 2 and 3 probabilities}
\label{FigB1}
\end{figure}

If you have $N$ numbers that are positive and add up to one, then all possible values of those numbers plot within a finite $N-1$ dimensional figure. That's because the locus of such points in $N$-dimensional space is the $N-1$ dimensional analog of a triangle (i.e. an $N-1$ simplex). For example, the composition of a binary solid-solution series can be represented by a straight line (1-simplex) that runs from (1,0) to (0,1). The composition of a ternary mixture in 3-space falls within a flat 2D triangle (2-simplex) with vertices at (1,0,0), (0,1,0) and (0,0,1). Figure \ref{FigB1} illustrates how each of these structures is embedded in the higher dimensional space of which it is a part. 

The 3D locus in four-dimensional space of a quaternary mixture is, similarly, a tetrahedron (3-simplex) with vertices at (1,0,0,0), (0,1,0,0), (0,0,1,0), and (0,0,0,1). Higher dimensional spaces work the same way, although they (and their embeddings) are harder to visualize. Probabilities as well as compositions are generally positive numbers that, taken together in a complete set, all sum to 1. Thus $N$ probabilities can also be mapped to the space within an $N-1$ simplex.

A useful feature of these simplex mappings is the ability to 
project to lower dimension simply by labeling one vertex as 
a sum of probabilities.  This works because the sum of a subset 
(say $m$) of $N$ normalized probabilities is simply another 
probability.  This sum, in turn, when combined with the 
($N-m$) non-subset probabilities, continues to add to one.  
While the original $n$ probabilities are represented by an 
($N-1$) simplex, the new ($N-m+1$) probabilities map to an 
($N-m$) simplex for easier visualization.  Of course, this
projection process also throws away information.  

To avoid this loss of information, one can use the 
sub-set sum (of $m$-probabilities above) as a ``dipole 
separation vertex''.  For the case when $N=6$ and $m=3$, 
each set of probabilities can then be represented 
by a pair of dots on parallel planes of fixed separation.  
The planes reside in vertex-to-vertex three 
simplexes, as shown is Fig. \ref{FigX}.  The choice 
of the dipole-separation subset (e.g. use of $f_{i1}+f_{i2}+f_{i3}$ 
versus $f_{i1}+f_{i3}+f_{i5}$) affects which patterns are 
highlighted.  This symmetry breaking, of course, 
may be an advantage if the system of probabilities 
being visualized has a matching asymmetry.


\bibliography{ifzx}

\end{document}